\documentstyle[epsfig,12pt]{article}
\newlength{\dinwidth}
\newlength{\dinmargin}
\newcommand{\ba}{\begin{array}}
\newcommand{\ea}{\end{array}}
\newcommand{\be}{\begin{equation}}
\newcommand{\ee}{\end{equation}}
\newcommand{\bea}{\begin{eqnarray}}
\newcommand{\eea}{\end{eqnarray}}

\def\parallel{| \hskip-0.03cm |}

\def\bee{\begin{eqnarray}}
\def\eee{\end{eqnarray}}
\def\be{\begin{equation}}
\def\ee{\end{equation}}

\begin{document}
\thispagestyle{empty}
\addtocounter{page}{-1}
\begin{flushright}
SNUTP 98-018\\
{\tt hep-th/9803135}\\
revised version\\
\end{flushright}
\vspace*{1.3cm}
\centerline{\Large \bf Wilson-Polyakov Loop at Finite Temperature}
\vspace*{0.3cm}
\centerline{\Large \bf in}
\vspace*{0.3cm}
\centerline{\Large \bf Large $N$ Gauge Theory and Anti-de Sitter 
Supergravity 
\footnote{
Work supported in part by KOSEF SRC-Program, Ministry of Education Grant 
BSRI 97-2418, SNU Faculty Research Fund and Korea Foundation for Advanced 
Studies Faculty Fellowship, by GIF - the German-Israeli Foundation
for Scientific Reseach and by the European Commission TMP Programme
ERBFMRX-CT96-0045.}}
\vspace*{1.2cm} \centerline{\large \bf Soo-Jong Rey${}^a$, 
Stefan Theisen${}^b$ and Jung-Tay Yee${}^a$}
\vspace*{0.8cm}
\centerline{\large\it Physics Department, Seoul National University,
Seoul 151-742, KOREA${}^a$}
\vspace*{.3cm}
\centerline{\large\it Sektion Physik, Universit\"at M\"unchen, D-80333,
M\"unchen, GERMANY${}^b$}
\vspace*{1.5cm}
\centerline{\Large\bf abstract}
\vspace*{0.5cm}
Aspects of $d=4, {\cal N}=4$ superconformal $U(N)$ gauge theory are studied at 
finite temperature. Utilizing dual description of large $N$ and strong 
coupling limit via Type IIB string theory compactification on Schwarzschild 
anti-de Sitter spacetime, we study correlations of Wilson-Polyakov loops and 
heavy quark potential thereof. We find that the heavy quark potential is 
Coulomb-like and has a finite range, as expected for gauge theory in high 
temperature, deconfinement phase. The potential exhibits finite temperature 
scaling consistent with underlying conformal invariance. We also study isolated
heavy quark on probe D3-brane world-volume and find supporting 
evidence that near
extremal D3-branes are located at Schwarzschild horizon.
\vspace*{1.1cm}

\baselineskip=18pt
\newpage
%
\section{Introduction}
\setcounter{equation}{0}

To understand the large $N$ and classical strong coupling behavior of 
$SU(N)$ gauge theories has been an open problem for decades~\cite{thooft, 
witten}. 
Recently, with better understanding of D-brane world-volume dynamics,
connections to string theory and therefore new approaches to the problem 
have become available.
Built on earlier studies of the near-horizon geometry of D- and M-branes
and their absorption and Hawking emission processes~\cite{klebanov},
Maldacena has put forward a proposal~\cite{maldacena} 
for the large $N$ limit of world-volume quantum field theories. 
In particular, the gauge theory Green functions can be calculated
via S-matrix elements of anti-de Sitter supergravity~\cite{klebanov2,
witten2}.

Among field theories studied so far, 
the most tractible case is $d=4, {\cal N}=4$ 
supersymmetric gauge theory with gauge group $SU(N)$ (and truncations to
${\cal N} = 2, 1, 0$ theories with vanishing beta functions)~\cite{adsnoise}.
The theory is superconformally invariant and is realized as the 
world-volume theory of $N$ coincident D3-branes in
Type IIB string theory. The latter configuration induces the near-horizon 
geometry of $adS_5 \times S_5$ 
\be
ds_{\rm extreme}^2
= \alpha' \left[ \,
{ {\rm U}^2 \over g_{\rm eff} } \left( -dt^2 + d {\bf x}_{\parallel}^2 \right)
+ {g_{\rm eff} \over {\rm U}^2} d {\rm U}^2 
+ g_{\rm eff} d {\bf \Omega_5}^2 \, \right]
\ee
whose radius of curvature-squared is 
$ g_{\rm eff} = g^2_{\rm YM} N$, where $g^2_{\rm YM} = \lambda_{\rm IIB}$
, and whose Freund-Rubin background is 
provided by self-dual flux $Q_5 = \oint_{S_5} H_5 = \oint_{S_5} 
{}^*H_5 = N$. Hence, in the large $N$ and strong coupling $g_{\rm eff} \gg 1$ 
limit, the gauge theory is dual to Type IIB supergravity on 
$adS_5 \times S_5$.

Maldacena's proposal is not restricted to zero temperature 
and can be extended to the study of thermodynamics at finite temperature.
At finite temperature, the large $N$ and strong
coupling limit of $d=4, {\cal N}=4$ supersymmetric gauge theory is dual to 
the near-horizon geometry of near extremal 
D3-branes~\cite{horowitzstrominger} in 
Type IIB string theory.
The latter is given by a Schwarzschild-anti-de Sitter Type IIB supergravity
compactification:
\be
ds^2 = \alpha' \left[
{1 \over \sqrt G} \left( - H dt^2 + d {\bf x}_{\parallel}^2 \right)
+ \sqrt{G} \left( {1 \over H} d {\rm U}^2 + {\rm U}^2 d {\bf \Omega}_5^2
\right) \, \right]
\label{sugra}
\ee
where
\bee
G &\equiv& {g^2_{\rm eff} \over {\rm U}^4} 
\nonumber \\
H &\equiv& 1 - { {\rm U}_0^4 \over {\rm U}^4 } 
\hskip1cm \left({\rm U}_0^4 = {2^7 \pi^4 \over 3 } g^4_{\rm eff} \, 
{\mu \over N^2}
\right) \, .
\label{harmonicftn}
\eee
The parameter $\mu$ is interpreted as the free energy density on the near
extremal D3-brane, hence, $\mu \approx (4 \pi^2 /45) N^2 T^4$. In the 
field theory limit $\alpha' \rightarrow 0$, $\mu$ remains finite. In turn, 
the proper energy $ E_{\rm sugra} = \sqrt{g_{\rm eff} / \alpha'} \mu / {\rm U}$
and the dual description in terms of modes propagating in the above
supergravity background is expected to be a good approximation.  

At a finite critical temperature $T = T_c$ , pure $SU(N)$ gauge theory 
exhibits a deconfinement phase transition~\cite{polyakov, susskind}. 
The relevant order parameter is the Wilson-Polyakov loop:
\be
P({\bf x}) = {1 \over N} {\rm Tr} {\cal P} \exp \left( i \int_0^{1 \over T} 
A_0 ({\bf x}) dt \right).
\label{polyakov}
\ee
Below the critical temperature $T < T_c$, $\langle P \rangle = 0$ and
QCD confines. Above  $T > T_c$, $\langle P \rangle$ is nonzero and takes
values in ${\bf Z}_N$, the center group of $SU(N)$~\footnote{
For review of gauge 
theory at finite temperature, see, for example, 
Refs.~\cite{gross, svetitsky}.}. 
Likewise, the two-point correlation of parallel Wilson-Polyakov loops 
\be
\Gamma({\bf d}, T) \quad \equiv \quad \langle P^\dagger({\bf 0}) 
\,  P ({\bf d}) \rangle_T
\quad = \quad e^{- {\cal F}({\bf d},T)/T } \,\, \approx 
\,\, e^{- V_{\rm Q {\overline Q}} ({\bf d}, T)/T }
\ee 
measures the static potential at finite temperature between quark and
anti-quark separated by a distance $d$.

At sufficiently high temperature, thermal excitations produce a plasma 
of quarks and gluons and gives rise to Debye mass $m_{\rm E} \approx
g_{\rm eff} T$ (which is responsible for screening color electric flux) and magnetic mass 
$m_{\rm M} \approx g^2_{\rm eff} T$ (which corresponds to the glueball mass
gap in the confining three-dimensional pure gauge theory). Their effects 
are captured by the asymptotic behavior of the heavy quark 
potential~\cite{braaten}:
\bee
\begin{array}{llll}
V_{\rm Q {\overline Q}} ({\bf d}, T)
\,\,\, \approx & - \, C_{\rm E} \, 
{1 \over |{\bf d}|^2} \,e^{- 2 \, m_{\rm E} |{\bf d}|} 
\,\, + \cdots & \qquad \qquad \quad & C_{\rm E} = {\cal O}(g^4_{\rm YM})\\
&&& \\
 & -  C_{\rm M} \, 
{1 \over |{\bf d}|} \, e^{- m_{\rm M} |{\bf d}|}
\,\, + \cdots &\qquad \qquad \quad & C_{\rm M} = {\cal O}(g^{12}_{\rm YM})
.  \end{array}
\label{debye}
\eee
In this paper, utilizing the aforementioned correspondence, 
we study correlators of Wilson-Polyakov loops and heavy quark potential 
at finite temperature in $d=4, {\cal N}=4$ supersymmetric gauge theory. 
At zero temperature,
heavy quark potential has been studied recently~\cite{reyyee, maldacena2} 
and Coulomb-type behavior consistent with the underlying conformal invariance
has been observed~\footnote{Quark-monopole potential has also 
been examined~\cite{minahan} utilizing 
triple string junction~\cite{triplestring}. }.
We will find that, at finite temperature, the heavy quark potential exhibits  
short-range asymptotic behavior as in Eq.~(\ref{debye}), but again in a manner 
fully respecting the underyling conformal invariance. 
We will also study an isolated 
single quark excitation on a probe D3-brane and find an indication that 
the near extremal D3-branes cannot reside behind the Schwarzschild horizon.
This provides a further support to the Maldacena's earlier 
observation~\cite{maldaprobe} that near-extremal D3-branes are located at 
the horizon.  

Concurrent with the first version of this paper, a preprint by E. 
Witten~\cite{wittennew} and a preprint by Brandhuber, Itzhaki, Sonnenschein 
and Yankielowicz~\cite{BISY} have appeared. 
The general gesults of Witten on the 
impossiblity of having a phase-transition at finite temperature for the 
$N=4$ theory led us to reconsider related speculations in the original version
of our paper. This also eliminated discrepancies in interpretation with the 
second paper mentioned.

\section{Static Quark Potential at Finite Temperature}
We want to study the dynamics of a test Type IIB string which ends on
a near-extremal D3-brane. Denote the string coordinates by $X^\mu(\sigma,
\tau)$, where $\sigma, \tau$ parametrize the string worldsheet. Low-energy
dynamics of the test string may be described via its
Nambu-Goto action, whose Lagrangian is~\footnote{
Throughout the paper, unless explicitly specified, we suppress the string 
tension $1/ 4 \pi \alpha' = 1$.} 
\be
L_{\rm NG} = -  
\int d\sigma
\sqrt{- {\rm det} h_{ab} } + L_{\rm boundary}.
\ee
Here, the induced metric on the worldsheet is 
\be
h_{ab} = G_{\mu \nu}(X) \partial_a X^\mu \partial_b X^\nu
\ee
To study the relevant string configurations of our interest, we take
$X^0 = t = \tau$ and decompose the nine spatial embedding  coordinates into
components parallel and perpendicular to the D3-branes:
\be
{\bf X} = \left({\bf X}_{\parallel}, \alpha' {\rm U}, \alpha'
{\rm U} {\bf \Omega}_5 \right) .
\ee
In the background metric Eq.(\ref{sugra})
straightforward calculation yields ($^{^.} \equiv \partial_t, \,\,
' \equiv \partial_\sigma$):
\bee
h_{00} &=& {1 \over \sqrt G} \left( - H + {\dot {\bf X} }^2_{\parallel} \right)
+ {\sqrt G} \left({1 \over H} {\dot {\rm U}}^2 
+ {\rm U}^2 {\dot {\bf \Omega} }_5^2 \right)
\nonumber \\
h_{11} &=& {1 \over \sqrt G} {{\bf X}'_{\parallel}}^2
+ {\sqrt G} \left( {1 \over H} {{\rm U}'}^2
+ {\rm U}^2 {{\bf \Omega}'_5}^2 \right)
\nonumber \\
h_{01} &=& {1 \over \sqrt G} {\dot {\bf X}}_{\parallel} \cdot
{\bf X}'_{\parallel}
+ {\sqrt G} \left( {1 \over H} \dot {\rm U} 
{\rm U}' + {\rm U}^2 \dot {\bf \Omega}_5 \cdot 
{\bf \Omega}'_5 \right)
\eee
{}From this, it is easy to see that a static configuration is described by
\be
L_{\rm NG} \rightarrow 
- \int d \sigma \sqrt{ {{\rm U}'^2+H {\rm U}^2{\bf \Omega}_5'^2} 
+ {H \over G} {{\bf X}'_{\parallel}}^2 }.
\label{nglagrangian}
\ee
Recalling that $G$ and $H$ are functions of $|{\bf X}_\perp| = \alpha'$U only,
the equations of motion are
\bee
\left( { {H\over G}{\bf X}'_{\parallel} \over 
\sqrt{ {\rm U}'^2+H {\rm U}^2{\bf \Omega}_5'^2 
+ {H \over G} {{\bf X}'_{\parallel}}^2 } } \right)' 
& = & 0
\nonumber \\
\left( { {H {\rm U}^2 {\bf \Omega}_5' \over 
\sqrt{ {\rm U}'^2+H U^2{\bf \Omega}_5'^2 
+ {H \over G} {{\bf X}'_{\parallel}}^2 }} } \right)' 
& = & 0
\nonumber \\
\left( { {\rm U}'\over 
\sqrt{ {\rm U}'^2+H {\rm U}^2{\bf \Omega}_5'^2 
+ {H \over G} {{\bf X}'_{\parallel}}^2 } } \right)' 
& = & {1\over 2}{{\bf X}_{\parallel}'^2\partial_{\rm U}
({H\over G})+{\bf \Omega}_5'^2 \partial_{\rm U} (H {\rm U}^2)\over 
\sqrt{ {\rm U}'^2+H {\rm U}^2{\bf \Omega}_5'^2 + {H \over G} 
{{\bf X}'_{\parallel}}^2 }}.  \eee

Following the prescription of Ref.~\cite{reyyee, maldacena2}, one now look 
for possible static string configurations that represent a pair of 
heavy quark and anti-quark. These strings are infinitely stretched out
of the $adS_5$ boundary and acts as static color sources of the boundary 
superconformal Yang-Mills theory. 
As in zero temperature situation~\cite{reyyee, maldacena2}, we will find 
that there are two possible string geodesics. The first is a pair of oppositely
oriented straight macroscopic strings. At finite temperature, a novelty
of this configuration
is that both strings end at $\alpha' {\rm U}_0$, not at $\alpha'$ U = 0. 
This is a consequence of Maldacena's observation~\cite{maldaprobe} that 
near extremal $N$ D3-branes are located at the Schwarzschild horizon.
The second is a U-shaped string inter-connecting quark and anti-quark.
The two possible configurations are illustrated in Figure 1.

\begin{figure}[t]
   \vspace{0cm}
   \epsfysize=8cm
   \epsfxsize=12.5cm
   \centerline{\epsffile{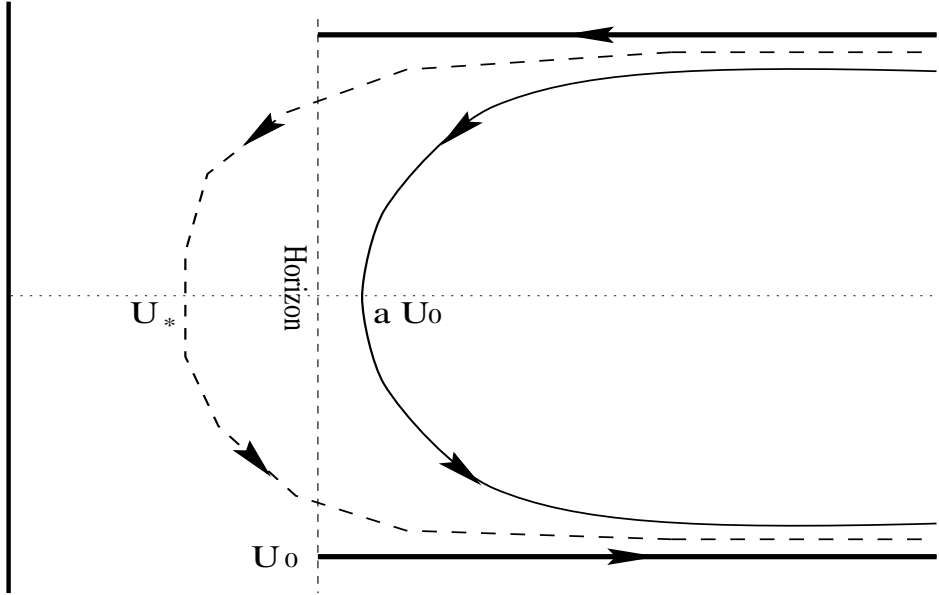}}
\caption{
\sl
Possible macroscopic string configurations representing various timelike
Polyakov-Wilson loops at finite temperature. The Schwarzschild horizon is
located at $\alpha' {\rm U}_0$. The first is a pair of straight, oppositely
oriented strings stretched between the $adS_5$ boundary and the Schwarzschild
horizon. The second is a U-shaped string runnig between quark and anti-quark.
The string tip is the closest point to the horizon and is located at
$a {\rm U}_0$. For comparison, U-shape string at zero temperature is also
shown.}
\end{figure}

We begin with the first configuration: a pair of straight macroscopic strings. 
The strings are stretched between $\alpha' {\rm U} = \infty$, the boundary of 
$adS_5$, and the Schwarzschild horizon, $\alpha' {\rm U}_0$:
\be
{\rm U} = \sigma, \qquad {\bf X}_{\parallel}={\rm constant},
\quad {\bf \Omega}_5 = {\rm constant}.
\label{radialstring}
\ee
Each macroscopic string of opposite orientation is a realization of 
timelike Wilson-Polyakov loops $\langle P \rangle$, $\langle P^\dagger
\rangle$ in Eq.~(4).
Likewise, two parallel macroscopic strings of opposite orientation 
(parametrized as in Eq.~(\ref{radialstring}) ) at a separation $|{\bf d}|$ 
is a realization of the two-point Wilson-Polyakov loop correlator 
$\langle P^\dagger({\bf 0}) \, P({\bf d}) \rangle_T$ in Eq.~(5).
As the two are oriented oppositely but have the same SO(6) angular position 
${\bf \Omega}_5$, the supersymmetry is completely broken and an attractive 
force will act between them. The force has no visible effect, however, as the 
quarks are heavy enough. In Figure 1, we have illustrated the configuration 
running along the $\alpha'$U-direction. Regularizing the string length
by $\Lambda = {\rm U} / {\rm U}_0$, the total energy of the quark pair $E_{\rm 
Q {\overline Q}}$ (measured in gauge theory unit) turns out equal to 
the mass of the quark pair at temperature $T$:
\be
E_{\rm Q {\overline Q}} (T, d) \quad = \quad 2 (\Lambda - 1) {\rm U}_0 = 
2 M_{\rm Q}; 
\qquad \qquad \qquad (\Lambda \rightarrow \infty).  
\ee

Of some interest is the fluctuation dynamics of the macroscopic strings.
The dynamics would be relevant for understanding Hawking radiation from
the Schwarzschild anti-de Sitter black hole Eq.~(2) along the string and 
the thermal 
distribution of string configurations. 
>From Eqs.~(7,10), one finds that harmonic fluctuations are governed by the
Lagrangian:
\be
L^{(2)} = {1 \over 2} \left[
\,\, \left( {1 \over H} {\dot {\bf X}_{\parallel}}^2 - {H \over G}
{{\bf X}'_{\parallel}}^2 \right)
+ \left( {G \over H} {\rm U}^2 {\dot {\bf \Omega}_5}^2
- H {\rm U}^2 {{\bf \Omega}_5'}^2 \right) \,\, \right] + \cdots.
\ee
Introducing Schwarzschild anti-de Sitter tortoise coordinate 
\be
r = \int d {\rm U} \, {{\sqrt G} \over H} \, ,
\ee
one finds that the equations of motion may be expressed as:
\be
\left[ \partial_t^2 - \partial_r^2 + M^2(r) \right] {\bf Y}_{\parallel} (r, t)
= 0
\ee
where $ {\bf Y}_{\parallel} (r, t) = {\rm U} {\bf X}_{\parallel} $ and
\be
M^2(r) = {2 \over g_{\rm eff}^2} {\rm U}^2 \left( 1 - {{\rm U}_0^8 
\over {\rm U}^8} \right) ,
\ee
and
\be
\left[ \partial_t^2 - \partial_r^2 \right] {\bf \Omega}_5 (r, t) = 0.
\ee
Fluctuations of the transverse field ${\bf Y}_{\parallel}$ obey 
the Klein-Gordon equation with spatially varying mass $M(r)$. 
Outside the Schwarzschild horizon, $M(r)$ is always positive
and grows indefinitely as the boundary of $adS_5$ is approached.
As this renders normalizable solutions, harmonic fluctuations along
the ${\bf X}_{\parallel}$ direction remain always bounded and stable.
Fluctuations of the angular field ${\bf \Omega}_5$ are free, hence, form
a $(1+1)$-dimensional Bose gas at Hawking temperature 
$T_H = {\rm U}_0/\pi g_{\rm eff}$. Logarithmic infrared divergence of the
latter system will then cause the classical string configuration to 
wander on $S_5$. Over a time interval $\Delta \tau$, for example, 
thermal ensemble average yields
\be
\langle\!\!\langle {\bf \Omega}_5^2 \rangle\!\!\rangle_{T} 
\quad
\approx \quad 5 \, {T_H \over 2 \pi } \cdot \Delta \tau .
\ee
At large $N$ and strong coupling $g_{\rm eff} \rightarrow \infty$, 
$M(r)$ vanishes and ${\bf Y}_{\parallel}$ fields
exhibit thermal infrared divergence as well. In turn, fluctuation of
${\bf X}_{\parallel}$ field is estimated
\be
\langle\!\!\langle {\bf X}_{\parallel}^2 \rangle\!\!\rangle_{T}
\quad
\approx \quad 
3 \, { 1 \over {\rm U}^2} {T_H \over 2 \pi } \cdot \Delta \tau .
\ee
The fluctuation is most pronounced near the Schwarzschild horizon. 

We now turn to the second configuration: a macroscopic U-shape string 
whose each end is connected to quark and anti-quark at the boundary of
$adS_5$. Parametrizing the string via static gauge: 
\be
{\bf X}_{\rm \parallel} = \sigma \hat{\bf n},\qquad {\bf\Omega}_5={\rm 
constant},
\label{orientation}
\ee
the equations of motion 
\bee
\left({{H\over G}\over\sqrt{{\rm U}'^2+{H\over G}}}\right)' & = & 0 
\nonumber\\
\left({{\rm U}'\over\sqrt{{\rm U}'^2+{H\over G}}}\right)'
& = & {\partial_{\rm U} ({H\over G})\over\sqrt{{\rm U}'^2+{H\over G}}}.
\eee
are solved straightforwardly by the first integral
\be
\left({G \over H} \right)^2
{\rm U}^{'2}
+ {G \over H} = {\rm constant}.
\label{firstintegral}
\ee
The left-hand side is greater than or equal to $- g^2_{\rm eff} / {\rm U}_0^4$. 
>From Eq.~(\ref{nglagrangian}), however, physically meaningful solution results
only when the first-integral is positive definite. Therefore, we take 
\be
\left({G \over H} \right)^2
{\rm U}^{'2}
+ {G \over H} = \left({ G \over H} \right)_* \equiv
{g_{\rm eff}^2 \over U_0^4 } {1 \over a^4 - 1},
\label{positiveintegral}
\ee
where the positive first integral is parametrized in terms of
$a \equiv {\rm U}_*/ {\rm U}_0 $~\footnote{The zero-temperature limit 
(which is also the extreme D3-brane limit) is reached by taking 
${\rm U}_0 \rightarrow 0$, $a \rightarrow \infty$ while 
holding ${\rm U}_0^4 (a^4 - 1) $ fixed. Also, positivity of 
$({G\over H})^2 {\rm U}'^2$ restricts the range of 
$a$ to $1\leq a\leq {\rm U}/ {\rm U}_0 < \infty$.}. 
One may interpret Eq.~(\ref{positiveintegral}) as a conserved `energy'
of a one-dimensional analog particle for which ${}'$ is treated as a time
derivative.
Simple phase-space consideration show that the entire configuration of 
the string lies outside the Schwarzschild horizon, i.e. ${\rm U} \ge
{\rm U}_0$. 

>From the first-integral Eq.~(\ref{positiveintegral}) one obtaines an  
implicit solution:
\be
{{\rm U}_0 \over g_{\rm eff} } \left( x - {d \over 2} \right) 
= \pm \sqrt{a^4 - 1} \int_a^Y \, 
{dy  \over \sqrt { (y^4 - 1) (y^4 - a^4) } },
\ee
where $Y \equiv {\rm U} / {\rm U}_0$ and ranges over $1 \le a \le Y$. 
Inside the square-root of the integral is a fourth-order polynomial in 
$y^2$. Such an integral can be always brought into a form of an elliptic 
integral by a change of variable $ \omega(y) = {1 \over 2} (y^2/a + a/y^2)$: 
\be
{ {\rm U}_0 \over g_{\rm eff} } \left( x - {d \over 2} \right) 
= \pm {1 \over 4 \sqrt 2}    {\sqrt{a^4 - 1} \over \sqrt {a^3} }
\left[
\int_\gamma^z {d \omega \over \sqrt { (\omega - 1) ( \omega^2 - \gamma^2)} } 
- \int_\gamma^z {d \omega \over \sqrt{(\omega + 1) (\omega^2 - \gamma^2)} } 
\right] \, .
\ee
Here, $z \equiv \omega(Y)$ and $\gamma \equiv \omega(a)$. Note that
$1 \le \gamma \le z$.
Explicitly, in terms of elliptic integrals,
\be
{ {\rm U}_0 \over g_{\rm eff} } \left(x - {d \over 2} \right) = 
\pm {1 \over 4 \sqrt \gamma } {\sqrt{a^4 - 1} \over \sqrt {a^3}}
\left[ {F} \left( {\rm sin}^{-1} \sqrt{ {z - \gamma \over z - 1}} , 
\sqrt{ \gamma + 1 \over 2 \gamma} \right)
 - {F} \left( {\rm sin}^{-1} \sqrt{z - \gamma \over z +1},
\sqrt{\gamma-1\over 2\gamma} \right) \, \right],
\ee
where ${F}(\alpha, \beta)$ denotes an elliptic integral of the first kind.
As in the zero temperature case, the parameter $a{\rm U}_0 \equiv {\rm U}_*$
is the U-location for the tip of the U-shaped string. Typical configuration 
of the string is illustrated in Figure 1.

The inter-quark distance $d$ is determined by the parameter $a$: 
\be
{{\rm U}_0\over g_{\rm eff}}{d\over 2} = 
{ 1 \over 4 \sqrt{\gamma}} {\sqrt{a^4-1}\over \sqrt{a^3}}
\left[{\bf K}\left(\sqrt{\gamma+1\over 2\gamma}\right)-
{\bf K}\left(\sqrt{\gamma-1\over 2\gamma}\right)\right],
\label{outdistance}
\ee
where {\bf K} denotes the complete elliptic integral of the first kind. 
The functional relation between $a$ and $d$ is plotted in Figure 2.
The striking difference from zero the temperature limit, which is also
plotted in the figure, is the presence of a maximal separation distance, 
$d \le d_{\rm max}$, in case the quark pair is connected by the U-shape 
string.
For a fixed $d \le d_{\rm max}$, Figure 2 shows that there are two possible
U-shaped string configurations at two different values of $a$. 
Recalling that the tip of the U-shape string is located 
at $a {\rm U}_0 = {\rm U}_*$ (See Figure 1), 
one would expect that a configuration with larger value of $a$ parameter  
has shorter length, hence, is an energetically more favorable 
configuration. 

\begin{figure}[t]
   \vspace{0cm}
   \epsfysize=7cm
   \epsfxsize=12.5cm
   \centerline{\epsffile{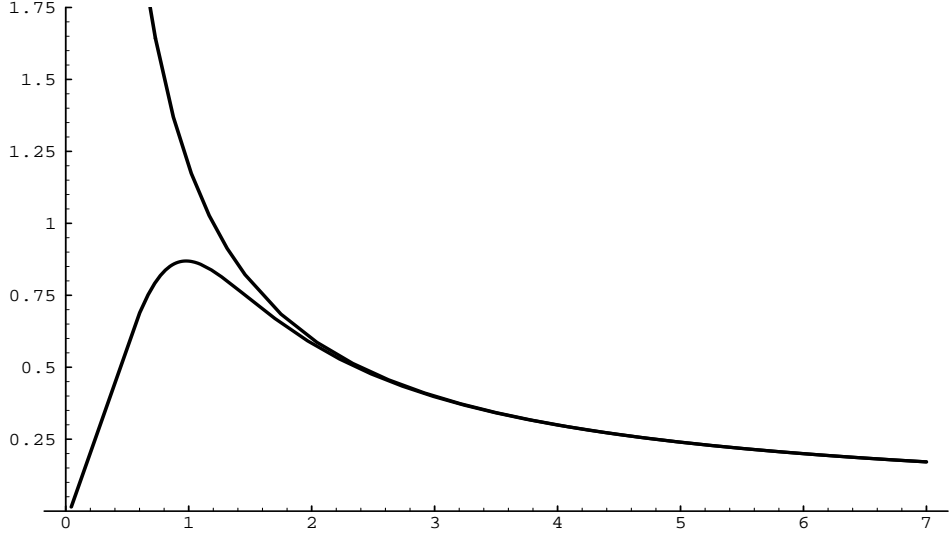}}
\caption{
\sl
Functional relation between the inter-quark distance $d$ and
$(a^4 - 1)^{1/4}$.
For $d \le d_{\rm max}$, the maximum inter-quark distance, 
there are two possible U-shape string configurations representing
the Wilson-Polyakov loop correlator at two different values of $a$. 
Also illustrated for comparison is the same functional relation at 
zero temperature. Note that, at any $d$, the zero temperature 
configuration is always unique. }
\end{figure}

The total energy of the U-shape string of inter-quark separation $d$ is 
\be
E_{\rm (Q {\overline Q})} (d, T) 
\quad =
\quad
\int_{- d/2}^{+d/2}d x\sqrt{{\rm U}^{'2}+{ H \over G}}
\,\, = \,\, 2{\rm U}_0\int^\Lambda_a d y\sqrt{y^4-1\over y^4-a^4},
\ee
which is again regularized by $\Lambda$ introduced in Eq.~(14).
Among the aforementioned possible string configurations,
the lower energy configuration will represent the stable Wilson-Polyakov 
loop correlator. For the U-shape string configuration, after subtracting
the mass of the quark pair, $2\, M_{\rm Q}$, the heavy quark potential
is:  
\bee
V_{\rm (Q {\overline Q})} ( d, T ) &\equiv&
E_{\rm (Q {\overline Q})} (d, T) \,\, - \,\, 
2 \, M_{\rm Q }
\nonumber \\
\nonumber \\
&=& 2{\rm U}_0\int^\Lambda_a d y\sqrt{y^4-1\over y^4-a^4}
-2{\rm U}_0(\Lambda-1).
\label{potentialenergy}
\eee
The dependence on the inter-quark distance $d$ is obtained by inverting
Eq.(29). The result is plotted in Figure 3.

In Figure 3, the energy of the U-shaped string configuration with the smaller
$a$-value is plotted as the upper dashed line 
while the one with the larger $a$-value is shown as the  
middle solid-dashed curve. The latter is clearly the lower energy 
configuration.
Its potential curve crosses zero at $d = d_*$.
Once $d \ge d_*$, Figure 3 shows that the configuration with two 
straight strings 
has a lower energy than the U-shape string configuration. 
As the two straight strings have no interaction energy, 
the heavy quark potential vanishes for $d \ge d_*$. 
Moreover, for all values of ${\rm U}_0$, one finds $d_* < d_{\rm max}$.
Summarizing, we have found that the two-point correlator 
of the Wilson-Polyakov 
loops Eq.~(5) in gauge theory is described in $adS_5$ supergravity by 
a U-shape string for $d < d_*$ and by a pair of straight strings
for $d > d_*$.    

\begin{figure}[t]
   \vspace{0cm}
   \epsfysize=7cm
   \epsfxsize=11.5cm
   \centerline{\epsffile{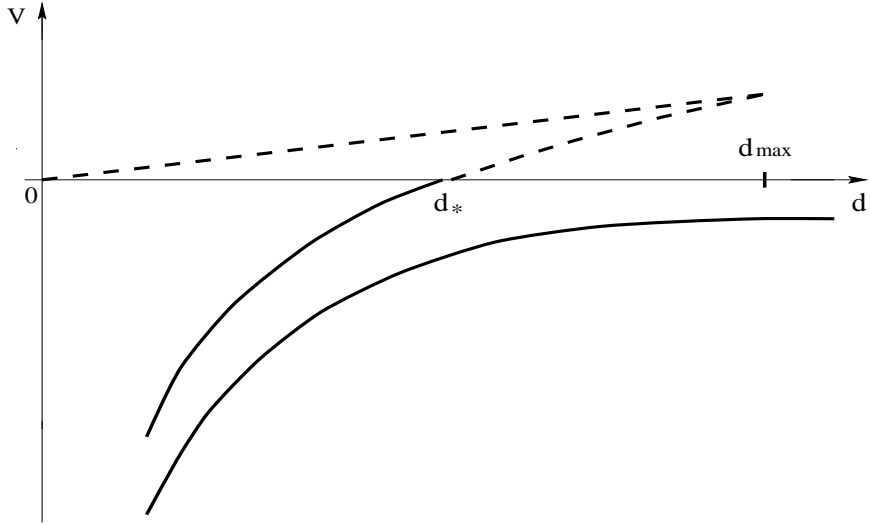}}
\caption{
\sl
The heavy quark potential $V_{\rm (Q {\overline Q})} (d)$ defined in Eq.~(31)
as a function of inter-quark distance $d$.
The upper dashed and the lower solid-dashed curves corresponds to two 
distinct string configurations for a given $d \le d_{\rm max}$ in Figure 2.
The dashed are U-shape string configurations that have higher total energy 
than the straight parallel string pair. Note that $d_* < d_{\rm max}$. 
For comparison, the heavy quark potential at zero temperature is shown 
in the lower solid curve.}
\end{figure}

We now derive an analytic expression for the heavy quark 
potential. At $d < d_*$, $a$ is large: $1 \le a \rightarrow
\infty$. Expanding Eq.~(\ref{outdistance}) in powers of $1/a$, one finds
\be
{ {\rm U}_0\over g_{\rm eff} }{d \over 2}=
c\left({1\over a}-{1\over 5 a^5}-{1\over 10 a^9}-\dots \right) \,\, ,
\label{expansion}
\ee
where
\be
c=\sqrt{2}{\bf E}(1/\sqrt{2})-{1\over\sqrt{2}}{\bf K}(1/\sqrt2)
={\sqrt{2}\pi^{3/2}\over\Gamma(1/4)^2}
\simeq 0.56 \,
\ee
and {\bf E} denotes the complete ellictic integral of the second kind.

Inverting eq.(\ref{expansion}), one obtains 
\be
a(d) = {2c g_{\rm eff}\over {\rm U}_0} {1 \over d}
-{1\over 5}\left({{\rm U} _0\over 2c g_{\rm eff}}\right)^3 d^3+\dots . 
\label{outasympto}
\ee
Likewise, expanding Eq.~(\ref{potentialenergy}) in powers of $1/a$ and
using the asymptotic relation Eq.~(\ref{outasympto}), 
one finally obtains asymptotic expansion of the heavy quark potential:
\bee
V_{\rm (Q{\overline Q})} (d, T) \,\, = \,\, 
\left\{ \begin{array}{lll}
 -4 c^2 g_{\rm eff} \left[ {1\over d}-{{\rm U}_0\over 2 c^2 g_{\rm eff}}
+{3\over 10}\left({{\rm U}_0\over 2 c g_{\rm eff}}\right)^4 d^3
+\dots\right]
& \qquad \qquad & ( d \le d_*) \\
&&\\
\qquad 0 \qquad \qquad & \qquad \qquad & (d \ge d_*).
\end{array} \right.
\label{outpot}
\eee
It is interesting to compare this with the pure SU(N) gauge theory, 
Eq.~(6).
Due to thermal excitations, both Eq.~(6) and Eq.~(\ref{outpot}) exhibit 
short-range interaction~\footnote{Recall that, in Eq.~(6), the electric mass 
arises perturbatively but the magnetic mass is a nonperturbative
effect.}. The two have different origin, however. The $g_{\rm YM}$ and $N$ 
dependence is strikingly different. The $1/|{\bf d}|$ dependence in 
Eq.~(6) is due to nonperturbative magneto-effects, 
whereas in Eq.~(\ref{outpot}) it is from the underlying conformal
invariance. The conformal invariance also constrains the heavy quark potential 
$V_{\rm (Q {\overline Q})}d$ to depend on temperature only through the 
unique scale-invariant combination $(T d)$. 
Another point to be emphasized is that, at any $d$, 
the heavy quark potential energy is always higher at finite temperature 
than at zero temperature (the lowest curve in Figure 3).


Actually there exists another string configuration that solves equations
of motion Eq.~(23). In fact, the first integral Eq.~(24) has a separatrix
at the Schwarzschild horizon ${\rm U} = {\rm U}_0$ and the extra solution
is a configuration inside the horizon. While its interpretation is not
clear, in view of recent claims~\cite{horowitzooguri} that large $N$ gauge 
theory also covers the Minkowski region behind the Schwarzschild 
horizon~\footnote{In fact, Minkowski spacetime formulation of finite 
temperature gauge theory is necessary if one were to study, for example,
transport phenomena.}, the new configuration might be relevant to issues 
such as resolution of black hole singularities~\cite{horowitzross}. 
Hence, we present briefly the configuration and leave its proper
interpretation for future work~\footnote{J. Minahan pointed out that
the solution presented in the first version of this paper did not 
have real energy in Minkowski spacetime. The solution actually describes an 
Euclidean tunnelling. He informed us that he has also found the configuration 
discussed below.}.
Solving Eq.~(\ref{positiveintegral}) for $0 \le {\rm U} \le
{\rm U}_0$,
\bee
{{\rm U}_0 \over g_{\rm eff} } \left( x - {d \over 2 } \right)
&=& \pm \sqrt{a^4 -1} \int_a^Y
d y { 1 \over \sqrt{(a^4 - y^4) ( 1 - y^4)} } ; \qquad \qquad
( 0 \le Y \le 1)
\nonumber \\
&=&
\pm {1 \over 4 \sqrt{\gamma}} {\sqrt{a^4 - 1} \over \sqrt{a^3}}
\left[ {F} \left({\rm sin}^{-1}
{\sqrt{2\gamma} \over \sqrt{z + \gamma}}
, {\sqrt{ \gamma + 1 \over 2 \gamma}} \right)
+ {F} \left({\rm sin}^{-1} {\sqrt{2 \gamma \over z + \gamma}}
, { \sqrt{\gamma - 1 \over 2 \gamma}} \right) \right],
\nonumber \\
\nonumber
\eee
where $\gamma = \omega (a) = {1 \over 2} (a + 1/a)$ and
$z \equiv \omega(Y)$. 
The distance $d$ between two ends of the macroscopic string at U = 0
is given by:
\bee
{ {\rm U}_0 \over g_{\rm eff}} {d \over 2}
= {1 \over 4 \sqrt \gamma}{\sqrt{a^4 - 1} \over \sqrt{a^3}}
\left[
{\bf K}\left(\sqrt{\gamma + 1 \over 2 \gamma} \right) + 
{\bf K}\left(\sqrt{\gamma - 1 \over 2 \gamma} \right) \right].
\nonumber
\eee
As the string extends only inside the horizon, the total energy:
\bee
E (a, T) 
= 2 {\rm U}_0 \int_0^1 dy \sqrt{\frac{1-y^4}{a^4 - y^4}}
\nonumber
\eee
is always finite.
In Figure 4, we have plotted the string configuration,
$d$ versus $a$-parameter, and the energy $E$ as a function of distance $d$ .

\begin{figure}[t]
   \vspace{0cm}
   \epsfysize=5cm
   \epsfxsize=15.5cm
   \centerline{\epsffile{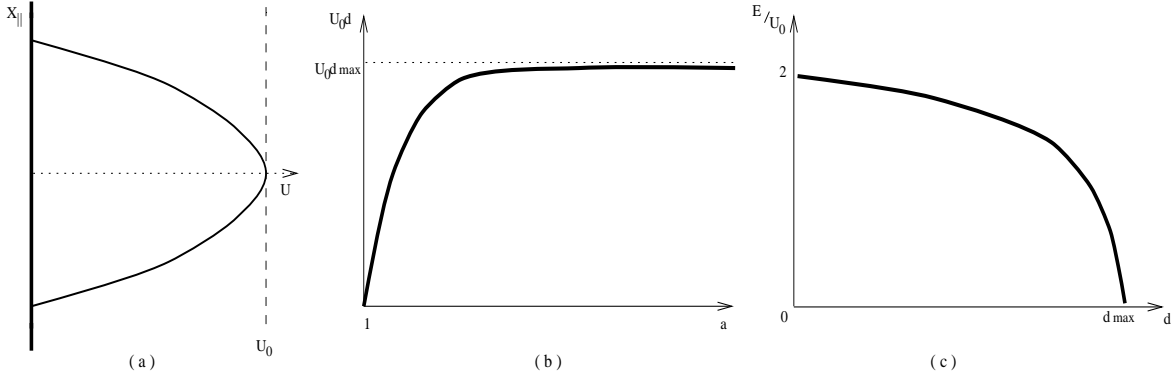}}
\caption{
\sl
(a) string inside Schwarzschild horizon, (b) distance $d$ versus first
integral parameter $a$, (c) total energy as a function of $d$. Note
the existence of a maximal distance $d_{\rm max}$. }
\end{figure}

\section{Near-Extremal Dynamics of Born-Infeld Quark}
In deducing that the Polyakov-Wilson loop correlator exhibits the 
expected deconfinement phase behavior at finite temperature, we have taken 
the interpretation that straight macroscopic strings terminate at the 
Schwarzschild horizon ${\rm U}_0$. 
This is a consequence of Maldacena's observation \cite{maldaprobe}
that near extremal D-branes are located at the horizon, not at the
singularity U = 0~\footnote{This point has also been stressed in 
~\cite{BISY}.}.
As this phenomenon has played a crucial role in deducing the behavior 
of the heavy quark potential, we now would like to examine it 
from a different angle to gain a better understanding. 

Introduce a probe D3-brane in the background of $N$ near extremal D3-branes.
World-volume dynamics of the probe D3-brane should be sensitive to the 
position of the N near extremal D3-branes and tell us where they are 
actually located.
To identify that, we consider a single quark state on the probe D3-brane
and study its fate as the probe D3-brane passes through the Schwarzschild
horizon. Unlike quark anti-quark pair, a single quark on a D3-brane carries 
net color electric flux. At weak coupling $\lambda_{\rm IIB} \rightarrow 0$, 
a single quark state is where Type IIB string is attached and the 
electric flux is conserved charge of the fundamental string
number~\cite{wittenbound}. We will now show that the electric flux 
of an isolated 
quark becomes unphysical once the probe D3-brane falls inside the 
Schwarzschild horizon. As this
implies that conserved string number becomes unphysical as well,  
we would conclude that fundamental string cannot penetrate through
the horizon.

At strong coupling $g_{\rm eff} \rightarrow \infty$, isolated quark
excitations are accurately described by the 
large $N$ resummed quantum Born-Infeld Lagrangian:
\be
L_{\rm DBI} = -{1 \over \lambda_{\rm IIB}}
\int d^3 x \, \sqrt{-{\rm det} \left( G_{ab} + F_{ab} \right) } - L_{\rm WZ}
\label{quantumdbi}
\ee
where $G_{ab} = \partial_a X^M \partial_b X^\nu G_{\mu \nu}
(X)$ denotes the induced metric and $F_{ab}$ is the worldsheet gauge field.
The Wess-Zumino term $L_{\rm WZ}$ provides a volume-dependent self-energy
potential.
{}From the Schwarzschild anti-de Sitter background geometry Eq.~(\ref{sugra}),
the quantum Born-Infeld Lagrangian for a probe D3-brane
is derived straightforwardly. For spherically symmetric excitations of
U and electric fields only
(${}' \equiv \partial_r, {\rm E} \equiv {\hat {\bf r}} \cdot {\bf E}$):
\be
L_{\rm DBI} =  -{4 \pi \over \lambda_{\rm IIB} }
\int r^2 dr \, {1 \over G} \left[
\sqrt{ H + G {{\rm U}'}^2 - {G \over H} {\dot {\rm U}}^2 - G {\rm E}^2 }
- 1 \right].
\label{dbi}
\ee
Canonically conjugate momenta of U and the gauge field are:
\bee
\lambda_{\rm IIB} \, \Pi_{\rm U}
&=& {\dot {\rm U}} {1/H \over \sqrt{ H + G {{\rm U}'}^2
- {G \over H} {\dot {\rm U}}^2
- G {\rm E}^2 }}
\nonumber \\
\lambda_{\rm IIB} \, \Pi_{\rm A}
&=& {\rm E} {1 \over \sqrt{H + G {{\rm U}'}^2
-{G \over H} {\dot {\rm U}}^2 - G {\rm E}^2 }}.
\eee
Thus, restricting to a static configuration, the equations of motion are
\bee
{1 \over r^2} \left( r^2 \Pi_{\rm A} \right)' &=& 0
\nonumber \\
{1 \over r^2} \left({ r^2 {\rm U}' \over \sqrt{ H + G {{\rm U}'}^2 - G {\rm E}^2} }
\right)' &=&
\partial_{\rm U} \left[ {1 \over G} \left(
\sqrt{H + G{{\rm U}'}^2 - G {\rm E}^2} - 1 \right) \right].
\label{eqns}
\eee
The first equation is nothing but the Gauss' law constraint on the  
world-volume. At zero temperature, $H \rightarrow 1$, and the two equations 
are solved simultaneously by a first-order BPS equation 
$E = {\rm U}'$~\cite{reyyee}.

At finite temperature, it is quite difficult to find an exact solution to
the equations of motion. Hence, we will consider approximate solutions
of an isolated quark, from which useful conclusions can still be drawn. 
Consider first the limit in which the electric field E is excited far 
more than the U-field gradient. From the second of Eq.~(\ref{eqns}), 
this is a good approximation in so far as U is not too small. We thus
set ${\rm U} = {\rm U}_p$ constant and solve the Gauss' law constraint.
The solution is
\be
E(r) = \sqrt{H} {Q_e \over \sqrt{r^4 + r_0^4} },
\label{electric}
\ee
where $r_0 = (G Q_e^2)^{1/4}$ and $H = H({\rm U}_p), G = G({\rm U}_p)$.
The solution Eq.~(\ref{electric}) represents an isolated quark of total
electric charge $Q_e$. The Schwarzschild metric function $H$ plays the 
role of a dielectric constant and screens the bare electric charge $Q_e$. 
The behavior of $H$, however, depends sensitively on the location ${\rm U}_p$
of the probe D3-brane. Outside the horizon ${\rm U}_p > {\rm U}_0$, 
$H > 0$ summarizes effects of thermal excitations. Inside the horizon,
${\rm U}_p < {\rm U}_0$, $H < 0$ and the electric field becomes purely 
imaginary. Moreover, Eq.~(\ref{dbi}) shows that the energy also becomes 
purely imaginary. Such pathological behavior is evaded only if D3-branes 
does not fall inside and stay exactly at the Schwarzschild horizon. 
Likewise, fundamental strings cannot penetrate through and would end at the horizon too.

The field energy of the isolated quark Eq.~(40) is calculated from the 
Lagrangian Eq.~(\ref{quantumdbi}):
\bee
E_{\rm Q} &=& {4 \pi \over \lambda_{\rm IIB}}
\int r^2 dr {1 \over G} \left[ \sqrt{H - G {\rm E}^2} - 1 \right]
\nonumber \\
&=& {4 \pi \over \lambda_{\rm IIB}}
{\sqrt{H} \over G} r_0^3
\int_0^\infty x^2 d x \left[ \sqrt{1 - {1 \over 1 + x^4} } - {1 \over
\sqrt H} \right].
\label{quarkenergy}
\eee
At the core $x \sim 0$, the integral is completely convergent. As $r_0$
depends on the function $G$, one may interpret the ultraviolet finiteness 
as a consequence of anti-de Sitter background geometry. On the other hand,
since $H < 1$, the quark energy is infrared divergent at $x \rightarrow
\infty$. The infrared divergence is absent at zero temperature, $H = 1$.
Hence, one may attribute it as a consequence of strong infrared divergence 
in gauge theory at finite temperature~\cite{gross}.
After regularizing the infrared divergence by replacing the second term 
in Eq.~(\ref{quarkenergy}) by unity, one finds that 
\be
E^{\rm reg}_{\rm Q} := {3 N \over \lambda_{\rm IIB}} {\sqrt{H} \over G}
\left({4 \pi \over 3} r_0^3 \right)
\hskip1cm N \approx 0.618.
\ee
The field energy clearly displays the fact that color charge is distributed
over a ball of radius $r_0$. The field energy on the probe D3-brane vanishes 
as the brane approaches the Schwarzschild horizon. This is a behavior 
consistent with Eq.~(14) that a pair of straight macroscopic string ending
on the Schwarzschild horizon has no interaction energy.

So far, we have approximated the single quark configuration by suppressing 
the U-field excitation. While exact considerationa are not feasible,
it is transparent what modifications one might expect for the 
isolated quark, Eq.~(\ref{electric}). Including the contribution of 
$G {\rm U}'^2$ to the Lagrangian Eq.~(\ref{dbi}), one recognizes 
that $\sqrt H$ in Eq.~(\ref{electric}) is now replaced by an implicit 
function $\sqrt{H + G {\rm U}'^2}$. The new color electric field is
physical only if ${\rm U}'^2 + H/G \ge 0$. Identifying $r = \sigma$,
one finds that this coincides precisely with the first-integral Eq.~(25).
Moreover, from Eq.~(\ref{eqns}), a real-valued U-field configuration is 
obtained only if ${\rm U}'^2 + H/G \ge {\rm E}^2 > 0$. Again, this results
in the condition of Eq.~(25). Thus, we conclude that D3-branes and 
fundamental strings ending on them cannot fall inside the Schwarzschild horizon.

\section{Discussions}
In this paper, we have studied the finite temperature behavior of
$d=4, {\cal N}=4$ superconformal Yang-Mills theory. Utilizing the
correspondence of the large $N$ and strong coupling limit to Type IIB 
supergravity compactification on Schwarzschild anti-de Sitter spacetime,
we have studied correlators of the Wilson-Polyakov loop and the heavy
quark potential. We have found that the potential exhibits behavior 
expected for the deconfinement phase at high temperature.

At zero 
temperature, the timelike Wilson loop is realized by a U-shaped macroscopic
string for any interquark separation. At finite temperature, we have seen
that the string configuration is either a pair of straight strings or
U-shaped. On the supergravity side, the new feature that emerges at finite 
temperature is the Schwarzschild black hole and its gravitational potential
in addition to the zero-temperature, anti-de Sitter gravity.
Hence, one may heuristically interpret that the transition from a U-shaped 
configuration to a pair of straight strings as the interquark distance
is increased is a result of pronounced gravitational attraction acting on 
the macroscopic string.

Crucial to the present study was the fact that macroscopic strings end at 
the Schwarzschild horizon, not at the singularity. To understand this 
behavior, we have studied a single quark state on a probe D3-brane in 
the Schwarzschild anti-de Sitter spacetime. We have found that the state
becomes unphysical if the probe D3-brane were to fall into the Schwarzschild
horizon. To avoid the pathology, D3-branes as well as fundamental strings 
ending on them are required to stay on or outside the horizon.

There remain many interesting issues to be explored. Viewing 
$d=4, {\cal N}=4$ superconformal gauge theory as a prototype quenched QCD, 
it would be most interesting to study in depth issues of confinement and 
chiral symmetry breaking. We also believe that possible resolutions of the 
black hole singularity in string theory might be intimately tied to
such issues in gauge theory.

We thank G. Horowitz, J. Minahan and S. Yankielowicz for comments on the 
first version of the paper. 
SJR and ST wish to thank M.R. Douglas, W. Lerche and H. Ooguri, organizers of 
{\sl Duality `98} workshop, and the Institute for Theoretical Physics for 
warm hospitality, where part of this work was accomplished.  
S.T. also acknowledges useful conversations with S. F\"orste.


\begin{thebibliography}{1}

\bibitem{thooft} G. `t Hooft, Nucl. Phys. {\bf B72} (1974) 461; {\sl 
Planar Diagram Field Theories}, in `Progress in Gauge Field Theory', NATO
Advanced Study Institute, eds. G. `t Hooft et.al. pp. 271-335 (Plenum, 
New York, 1984).

\bibitem{witten} E. Witten, Nucl. Phys. {\bf B149} (1979) 285;
Nucl. Phys. {\bf B160}(1979) 57. 

\bibitem{klebanov} I.R. Klebanov, Nucl. Phys. {\bf B496} (1997) 231;\\
S.S. Gubser, I.R. Klebanov and A. Tseytlin, Nucl. Phys. {\bf B499} (1997) 
217;\\
S.S. Gubser and I.R. Klebanov, Phys. Lett. {\bf B413} (1997) 41;\\
M.R. Douglas, J. Polchinski and A. Strominger, {\sl Probing Five-Dimensional
Black Holes with D-Branes}, {\tt hep-th/9703031}.

\bibitem{maldacena} J.M. Maldacena, {\sl The Large N Limit of 
Superconformal Field Theories and Supergravity}, {\tt hep-th/9711200}.

\bibitem{klebanov2} S.S. Gubser, I.R. Klebanov and A.M. Polyakov,
{\sl Gauge Theory Correlators from Noncritical String Theory},
{\tt hep-th/9802109}.

\bibitem{witten2} E. Witten, {\sl Anti-De Sitter Space and Holography},
{\tt hep-th/9802150}.

\bibitem{adsnoise} S. Ferrara and C. Fronsdal, {\sl Conformal Maxwell
Theory as a Singleton Field Theory on $AdS_5$, IIB Three-Branes and
Duality}, {\tt hep-th/9712239};\\
P. Claus, R. Kallosh, J. Kumar, P. Townsend and A. van Proeyen, 
{\sl Conformal Theory of M2, D3, M5 and D1+D5-branes}, {\tt hep-th/9801206};\\
N. Itzhaki, J. Maldacena, J. Sonnenschein and S. Yankielowicz, 
{\sl Supergravity and the Large N Limit of Theories with 16 Supercharges},
{\tt hep-th/9802042};\\
S. Ferrara and C. Fronsdal, {\sl Gauge Fields as Composite Boundary 
Excitations}, {\tt hep-th/9802126};\\
S. Ferrara, C. Fronsdal and A. Zaffaroni, {\sl On N=8 Supergravity on
$AdS_5$ and N=4 Superconformal Yang-Mills Theory}, {\tt hep-th/9802203};\\
G. Horowitz and H. Ooguri, {\sl Spectrum of Large N Gauge Theory from 
Supergravity}, {\tt hep-th/9802116};\\
S. Kachru and E. Silverstein, {\sl 4d Conformal Field Theories and Strings
on Orbifolds}, {\tt hep-th/9802183};\\
A. Lawrence, N. Nekrasov and C. Vafa, {\sl On Conformal Field Theories in
Four Dimensions}, {\tt hep-th/9803015};\\
O. Aharony, Y. Oz and Z. Yin, {\sl M-Theory on $AdS_p \times S^{11-p}$ and
Superconformal Field Theories}, {\tt hep-th/9803053};\\
S. Ferrara and A. Zaffaroni, {\sl N=1,2 4D Superconformal Field Theories
and Supergravity in $AdS_5$}, {\tt hep-th/9803060};\\
M. Bershadsky, Z. Kakushadze and C. Vafa, {\sl String Expansion as
Large N Expansion of Gauge Theories}, {\tt hep-th/9803076}.

\bibitem{horowitzstrominger}
G.T. Horowitz and A. Strominger, Nucl. Phys. {\bf B360} (1991) 197.

\bibitem{polyakov} A.M. Polyakov, Phys. Lett. {\bf 72B} (1978) 477.

\bibitem{susskind} L. Susskind, Phys. Rev. {\bf D20} (1979) 2610.

\bibitem{gross} D.J. Gross, R.D. Pisarski and L.G. Yaffe,
Rev. Mod. Phys. {\bf 53} (1981) 43.

\bibitem{svetitsky} B. Svetitsky, Phys. Rep. {\bf 132} (1986) 1.

\bibitem{braaten} E. Braaten and A. Nieto, Phys. Rev. Lett.
{\bf 74} (1995) 3530. 

\bibitem{reyyee} S.-J. Rey and J.-T. Yee, {\sl Macroscopic Strings as
Heavy Quarks in Large N Gauge Theories and Anti-de Sitter Supergravity},
{\tt hep-th/9803001}.

\bibitem{maldacena2} J. Maldacena, {\sl Wilson Loops in Large N Field 
Theories}, {\tt hep-th/9803002}.

\bibitem{minahan} J.A. Minahan, {\sl Quark-Monopole Potentials in Large
$N$ Yang-Mills}, {\tt hep-th/9803111}.

\bibitem{triplestring} O. Aharony, J. Sonnenschein and S. Yankielowicz,
Nucl. Phys. {\bf B474} (1996) 309;\\
J.H. Schwarz, Nucl. Phys. [Proc. Suppl.] {\bf 55B} (1997) 1;\\
K. Dasgupta and S. Mukhi, {\sl BPS Nature of 3-String Junctions},
{\tt hep-th/9711094};\\
A. Sen, {\sl String Network}, {\tt hep-th/9711130};\\
S.-J. Rey and J.-T. Yee, {\sl BPS Dynamics of Triple (p,q) String
Junction}, {\tt hep-th/9711202};\\
M. Krogh and S. Lee, {\sl String Network from M-Theory}, {\tt
hep-th/9712050};\\
Y. Matsuo and K. Okuyama, {\sl BPS Condition of String Junction from
M Theory}, {\tt hep-th/9712070};\\
O. Bergman, {\sl Three-Pronged Strings and 1/4 BPS States in N=4
Super-Yang-Mills}, {\tt hep-th/9712211};\\
M.R. Gaberdiel, T. Hauer and B. Zwiebach, {\sl Open String-String
Junction Transitions}, {\tt hep-th/9801205};\\
C.G. Callan Jr. and L. Thorlacius, {\sl Worldsheet Dynamics of
String Junctions}, {\tt hep-th/9803097}.

\bibitem{maldaprobe} J. Maldacena, Phys. Rev. {\bf D57} (1998) 3736.

\bibitem{wittennew} E. Witten, {\sl Anti-de Sitter Space, Thermal
Phase Transition and Confinement in Gauge Theories}, {\tt hep-th/9803131}.

\bibitem{BISY} A. Brandhuber, N. Itzhaki, J. Sonnenschein and
S. Yankielowicz, {\sl Wilson Loops in the Large N Limit at Finite
Temperature}, {\tt hep-th/9803137}.

\bibitem{horowitzooguri} G.T. Horowitz and H. Ooguri, {\sl Spectrum
of Large N Gauge Theory from Supergravity}, {\tt hep-th/9802116}.

\bibitem{horowitzross} G.T. Horowitz and S.F. Ross,
{\sl Possible Resolution of Black Hole Singularities from Large $N$
Gauge Theory}, {\tt hep-th/9803085}.

\bibitem{wittenbound} E. Witten, Nucl. Phys. {\bf B460} (1996) 335.


\end{thebibliography}
\end{document}